\documentclass[conference]{IEEEtran}
\IEEEoverridecommandlockouts
\usepackage{cite}
\usepackage{tabularx}
\usepackage{braket}
\usepackage{amsmath,amssymb,amsfonts}
\usepackage{graphicx}
\usepackage{textcomp}
\usepackage{xcolor}
\usepackage{subfigure}
\usepackage{subcaption}
\usepackage{soul} 
\usepackage{caption} 
\usepackage{multirow}
\usepackage{makecell}
\usepackage{algorithm}
\usepackage{algpseudocode}

\def\BibTeX{{\rm B\kern-.05em{\sc i\kern-.025em b}\kern-.08em
    T\kern-.1667em\lower.7ex\hbox{E}\kern-.125emX}}

\begin{document}

\title{
Watermarking of Quantum Circuits
}

\author{\IEEEauthorblockN{ Rupshali Roy}
\IEEEauthorblockA{\textit{School of EECS}\\
\textit{Penn State University,PA, USA}\\
rzr5509@psu.edu}
\and
\IEEEauthorblockN{Swaroop Ghosh}
\IEEEauthorblockA{\textit{School of EECS}\\ 
\textit{Penn State University,PA, USA}\\
szg212@psu.edu}
}
\maketitle
\begin{abstract}
Quantum circuits constitute Intellectual Property (IP) of the quantum developers and users, which needs to be protected from theft by adversarial agents e.g., the quantum cloud provider or a rogue adversary present in the cloud. This necessitates the exploration of low-overhead techniques applicable to near-term quantum devices, to trace the quantum circuits/algorithms’ IP and their output. We present two such lightweight watermarking techniques to prove ownership in the event of an adversary cloning the circuit design. For the first technique a rotation gate is placed on ancilla qubits combined with other gate (s) at the output of the circuit. For the second method, a set of random gates are inserted in the middle of the circuit followed by its inverse, separated from the circuit by a barrier. These models are combined and applied on benchmark circuits and the circuit depth, 2-qubit gate count, probability of successful trials (PST) and probabilistic proof of authorship (PPA) are compared against the state-of-the-art. The PST is reduced by a miniscule 0.53\% against the non-watermarked benchmarks and is up to 22.69\% higher compared to existing techniques. The circuit depth has been reduced by up to 27.7\% as against the state-of-the-art. The PPA is astronomically smaller than existing watermarks.

\end{abstract}


\section{Introduction}
Quantum computing is a rapidly growing field which could potentially revolutionize many industries, such as drug discovery, financial modeling, and material science. The principles of quantum mechanics give quantum bits or qubits the ability to exist in multiple states simultaneously, allowing quantum computers to solve complex problems exponentially faster than classical computers. 
The implementation of quantum algorithms on quantum computers has been realised using several different qubit technologies such as trapped ion qubits, superconducting qubits, quantum dots, photonic qubits and diamond nitrogen-vacancy centers . The potential advantages offered by quantum computing have inspired significant research efforts globally. Commercial quantum computers have already been brought into existence by companies like Microsoft, IBM and Google. However, in spite of all the potential, the technology is still taking its first steps, and there are many obstacles to be surmounted before it can be used widely.

A quantum circuit constitutes an ordered sequence of quantum gates that carry out specific operations on qubits to solve a problem. This task is complex and requires tools and domain knowledge. Quantum circuits/algorithms and their measured states are confidential information of the circuit designers, developers and users, i.e., it incorporates Intellectual Property (IP) that needs to be protected from theft by adversarial agents. 
However the current costs and queue wait time of trusted quantum services is high. This has motivated the rise of 3rd party quantum cloud providers who may charge considerably less and offer quantum hardware readily. Reliance on such cloud services may help in cutting down costs and computation time but can give rise to security breaches due to illegitimate cloning/theft of circuit designs by an adversary. The adversary may be an agent of the untrusted cloud provider, or an adversary who has gained unauthorised access to privileged information. This can lead to considerable financial loss for the researchers/organization that originally created the circuit and can also slow down the pace of the field of quantum computing. It is thus imperative to develop low-overhead techniques that trace the quantum algorithms’ IP and their output, which are applicable to near-term quantum devices. We present two lightweight watermarking techniques for quantum circuits to help the designer to prove ownership in a court of law in the event of an adversary stealing the design. 

{\textbf{Paper Organization:} Section II provides background on quantum computing and related works. We introduce the proposed watermarking models in Section III and discuss the methodology and results in Section IV. The security analysis and comparison against state-of-the-art watermarking models is covered in Section V. Finally, Section VI draws conclusion.}

\section{Background}
\subsection{Quantum Computation Preliminaries} 
\subsubsection{Qubits} {Qubits are the smallest units of quantum information. 
A two-level system is the most common way to represent a qubit, with the basis states $\ket{1}$ and $\ket{0}$. The state of a single qubit can be denoted by a linear combination of these two states, represented by $\psi$ = $\alpha$ $\ket{0}$ + $\beta$ $\ket{1}$, where $\alpha$ and $\beta$ are complex numbers and the squared magnitudes of $\alpha$ and $\beta$ denote the probabilities of the qubit being measured in the states $\ket{0}$ and $\ket{1}$, respectively and the sum of these magnitudes is equal to 1}.

\subsubsection{Quantum gates} Quantum gates are used to perform specific operations on qubits such as changing the state of a qubit, entangling multiple qubits, and creating superposition states.They are each represented by a unitary matrix which describes the operation that is performed on the quantum state and satisfies the condition $\mathbf{U}^\intercal$U= I, where $\mathbf{U}^\intercal$ refers to the conjugate transpose of U and I is the identity matrix. The rotation gates (Rx,Ry,Rz), Hadamard gate, Pauli-X gate, controlled-NOT (CNOT) gate are some of the commonly used quantum gates. Rotation gates Rx (or Ry) rotate the qubit state vector on the x-axis (or y-axis) of the Bloch sphere by a phase specified in radians. The Hadamard gate creates a superposition state with equal probability amplitudes of the $\ket{1}$ and $\ket{0}$ states. The Pauli-X gate flips the state of a qubit, and the CNOT gate (a two-qubit gate) flips the target qubit if the control qubit is in the state $\ket{1}$. 

\subsubsection{Basis gates and coupling constraints} {Quantum computers support a limited number of single and multi-qubit gates in practice, known as basis gates or native gates of the hardware. For instance, IBM quantum computers use the following native gates: CNOT (two-qubit), u1, u2, u3 and id (single-qubit). The quantum circuit may however contain complex gates that are not native to the hardware e.g., the Toffoli gate is not native to the IBM quantum computers. Hence before execution, such gates in a quantum circuit are decomposed into the basis gates. In addition, the two-qubit operation (CNOT) is permitted only between connected qubits. Such restrictions in two-qubit operations in any target hardware are called coupling constraints. Swap gates are inserted to meet the coupling constraints if a two qubit gate is applied between physically disconnected qubits.}

\subsubsection{Compilation} Quantum circuit compilers such as Qiskit apply operations (e.g., insert SWAP gates) to the input circuits to meet the hardware's coupling constraints. They also provide optimization of higher-level circuits using single/multi-qubit gate merging, cancellation, rotation and gate-reordering. Qiskit supports barriers between circuit partitions to restrict such optimizations across partitions.

\subsubsection{Quantum Approximate Optimization Algorithm (QAOA)} {This is a hybrid quantum-classical variational algorithm created to solve combinatorial optimization problems. It applies a parameterized quantum gate sequence iteratively to the qubits in a quantum circuit. QAOA is constituted by alternating layers of the Cost Hamiltonian (HC) and the Mixing Hamiltonian (HM) operations. The objective function of the optimization problem is encoded by the HC, while the HM introduces quantum fluctuations between different configurations.}
{Approximation ratio (AR) $\frac{C_{\text{QAOA}}}{C_{\text{max}}}$, i.e., the ratio of the cost associated with the solution output by QAOA to that of the optimal solution is a typical measure of the QAOA performance. Theoretically, AR increases with increasing layers $p$.}

\subsubsection{Total Variation Distance}{TVD is a measure of how much two probability distributions diverge from each other. Given two probability distributions \( A \) and \( B \) defined on the same sample space, the total variation distance between \( A \) and \( B \) is defined as: \text{TVD}(A, B) = \(\frac{1}{2} \sum_{x} |A(y) - B(y)|\)where the sum is taken over all possible outcomes \( y \) in the sample space.}

\subsubsection{Probability of Successful Trials (PST)} {PST of a quantum circuit is given by: \text{PST} = \(\text{num}_{\text{trials=initial}}/{{\text{num}_{\text{trial}}}}]\). \(num_{trials=initial}\) denotes the number of trials with an output identical to that of the initial state. num\_{trials}  represents the total number of trials.}

\subsection{Existing work on quantum circuit watermarking}
Some of the existing works on quantum circuit watermarking embed a secret signature in the form of additional gates or modified control parameters during the decomposition phase to verify the ownership of the IP \cite{b24}. However this approach involves adding gates to the main circuit for which the equivalent matrix is not necessarily unity, altering the functionality of the circuit. Another work transforms the input problem of QAOA, to make the resulting circuit and the outputs indecipherable to the server \cite{b25}. In \cite{b26}, the authors introduce multilevel watermarking by adding signatures at the decomposition, mapping and scheduling stages. However, such a scheme would not be universal as the adversary may use a different target hardware or simulator with a different coupling map and different native gate set to run the circuit. In such situations, the watermark will be modified and/or even removed. Our approach will eliminate such weaknesses since the added gate set for watermarking will be equivalent to a unity matrix or the rotation gates on the ancilla qubits in combination with another gate (CNOT in this work) will not disturb the base circuit functionality. 
\section{Attack Model and Proposed Watermarks}
\subsection{Attack model}
Quantum circuits are valuable IPs since it takes labor and time to create. 
We assume that the user may employ unreliable or less-trusted third-party quantum cloud providers due to the scarcity of affordable quantum cloud services with a reasonably short wait queue. However, using an untrusted cloud provider poses significant security risks due to possible illegitimate cloning/theft of circuit designs by an adversary. The objectives could be to (a) sell the designer's IP for direct financial profit, (b) pretend to be the original company designing the original circuitry, thereby hurting their brand and market share, (c) tamper with the counterfeit circuit or introduce hardware Trojans before circulating it in the supply chain. Such counterfeit copies can degrade the reliability of the computation or leak data. 
We assume that an unreliable third party hosts the cloud service remotely and a rogue adversary can break into the cloud storage, thus gaining unauthorized access to proprietary circuit designs. Another scenario is that the cloud provider itself could be an untrusted/rogue adversary or may possess an insider adversary who may retain illegitimate copies of designers' circuits stored in their system. 

\subsection{Rotation-gate based watermarking}
To protect the quantum circuit designers' IP, we propose an additive watermark design where a rotation gate (Rx or Ry) is added on an auxiliary/ancillary qubit. Since the watermark is applied on the ancilla qubit the functionality of the circuit is not affected. Prime candidates for such watermarking would include any circuits where the number of input qubits are larger than the number of output qubits. For example, the 4gt ‘x’ series of circuits take 4 qubits at the input, and outputs a ‘1’ if the number entered is larger than x, and 0 otherwise, thus having only 1 functional qubit at the output. However, they use many ancilla qubits. We will add rotation gate (s) at the output of one of the ancilla qubits, in addition to entangling with other ancilla qubits (CNOT in this paper) to create our watermark.


For circuits where all the qubits at the output are functional (i.e., no ancilla/auxiliary qubits), we can add an auxiliary qubit and entangle it with a functional qubit using a CNOT gate in order to ensure there is no functional impact for the purpose of watermarking. In this work we apply this concept on 3-node, 4-node and 5-node optimized QAOA circuits designed to solve maxcut problems as well.

We want the output probability vector of the watermarked circuit to be as far from the baseline circuit as possible, so that they may be distinguishable from non-watermarked circuit. This can be quantified in terms of TVD. Since users as well as adversaries may execute a given circuit on different backends, we aim to maximize the TVD between the probability distributions on the watermarked circuit when executed on different backends with respect to the baseline circuit which is executed on a fixed backend (in this case FakeValencia). To this effect, we find the phase of rotation that yields the maximum TVD over a range of benchmark circuits.
A watermark with specific angle of rotation would offer a low collision probability with genuine design (or a different watermark) by trusted designers, thus providing high ownership, quantified by the PPA (probabilistic proof of authorship). 

\subsection{Random gate insertion based watermarking}
In this method, we insert random gates in the circuit, followed by their inverse gates. The watermark thus created is separated from its inverse by a barrier to prevent optimization/cancellation. 
Since the selection of gates for the watermark is random, the probability of collision with a different designer's watermark is very low, establishing high ownership, as low as $2.4e^{-7}$ across the chosen benchmarks when applied in combination with the additive model described above. We choose a Pauli-X gate and a CNOT gate for our design.

\begin{figure*}
        \centering         
        \begin{minipage}{0.4\textwidth}
                \centering
                    \includegraphics[width=\columnwidth]{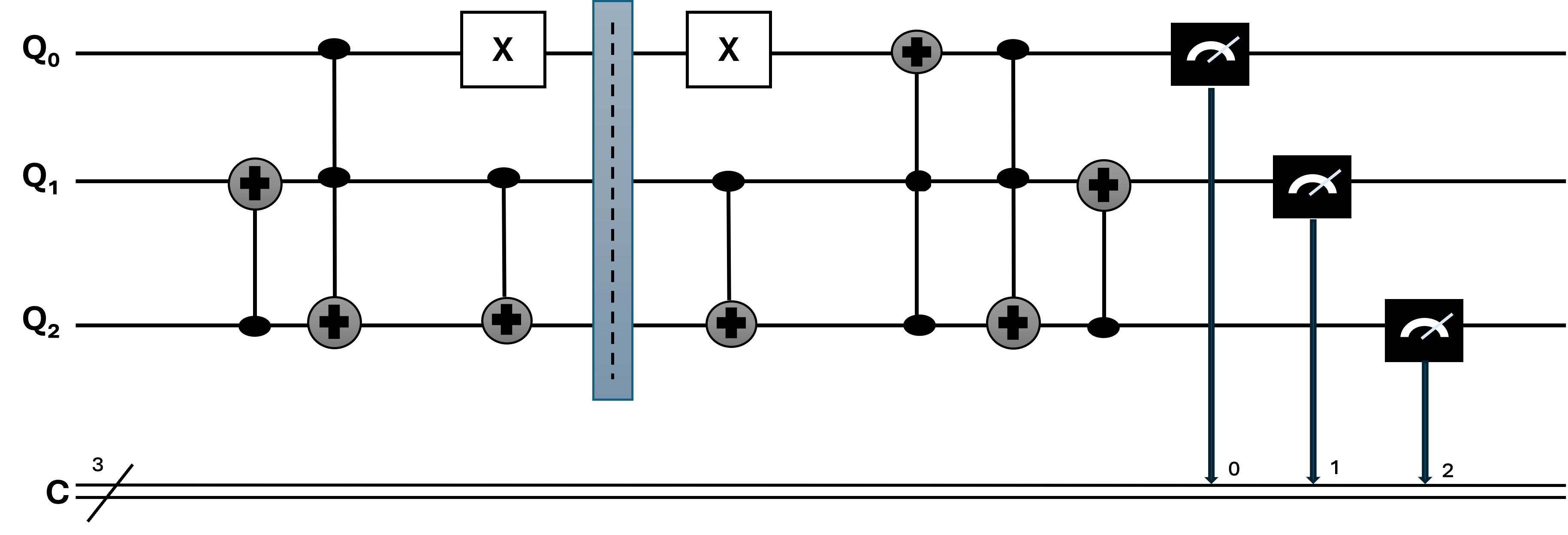}
                 \vspace{-5mm}
                 (a)
         \label{miller_wm}
        \end{minipage}%
        \begin{minipage}{0.4\textwidth}
                \centering
                    \includegraphics[width=\columnwidth]{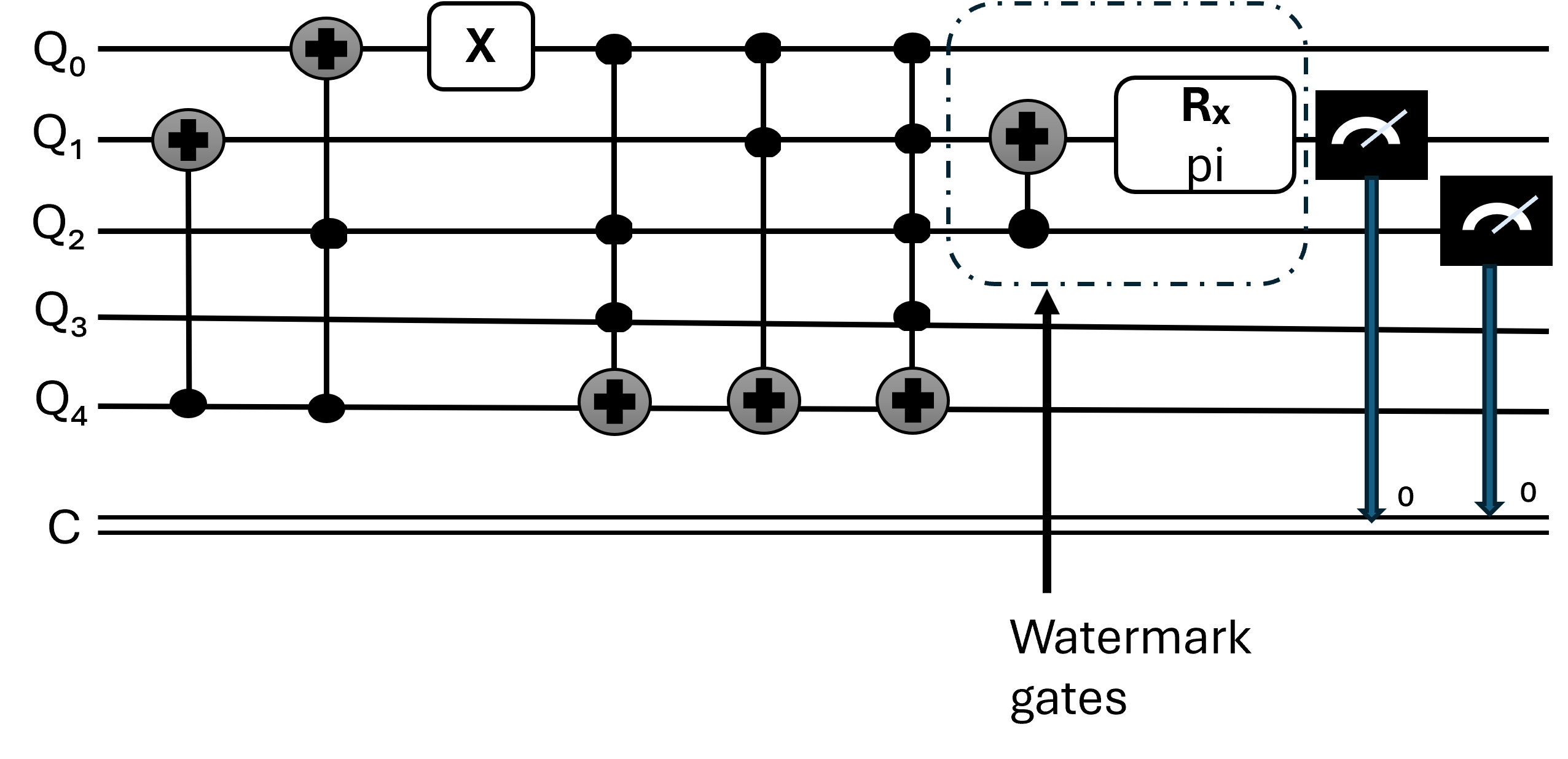}
                 \vspace{0mm}
                 (b)
        \label{4gt5_wm}
        \end{minipage}%
        \vspace{-1mm}
        \caption{(a) Random gate based watermarking applied on a Miller circuit, (b) rotation gate based watermarking applied on 4gt5 circuit. Q4 is the functional qubit while the others are ancillary. We choose Q1 \& Q2 to place our watermark. Other qubits can also be used for watermarking.}   
        \vspace{-6mm}

        \label{miller_wm_4gt5_wm}
\end{figure*}

\subsection{Retrieval of watermark}
\begin{figure}
    \centering
        \includegraphics[width=0.65\columnwidth]{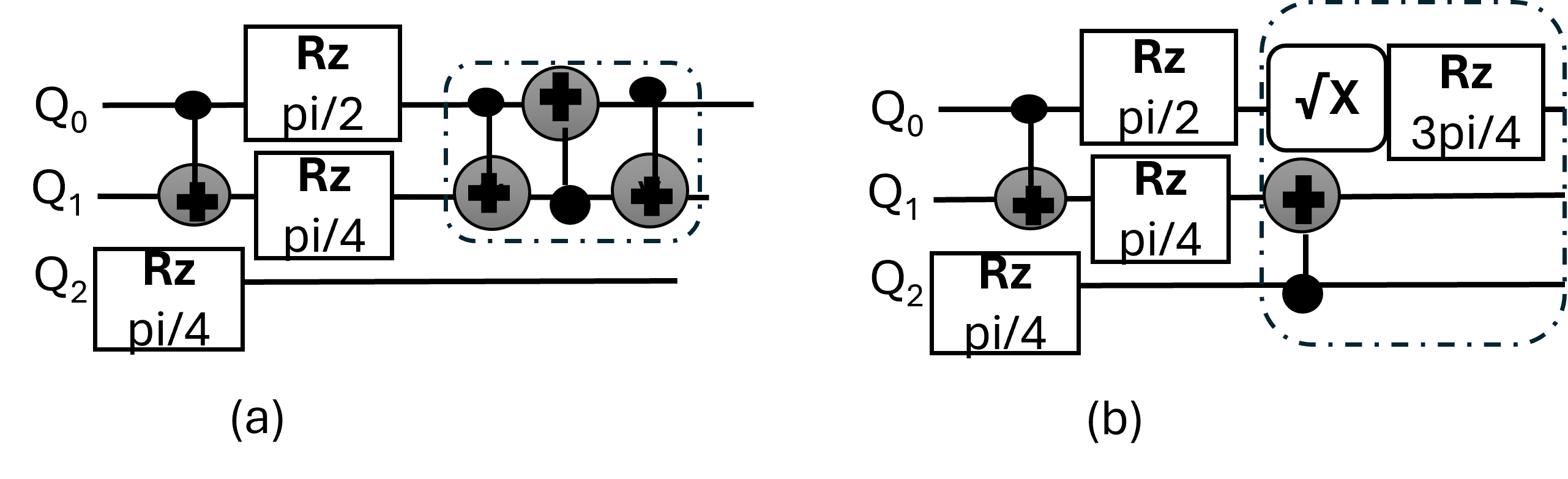}
        \caption{(a) Baseline circuit transpiled on FakeValencia and (b) watermarked circuit transpiled on FakeLagos.} 
        \label{retrieval}
     \vspace{-6mm}
\end{figure}

In the event of the circuit being cloned by an adversary, the designer may decide to take legal recourse. To prove ownership, the watermark needs to be retrieved from the circuit. We propose a heuristic (Algorithm-\ref{watermark}) to accomplishes this. The adversary and the designer may be using different quantum hardware to transpile the circuit. These may have different logical-physical qubit mappings and SWAP gate implementations to move the qubits around to fit their respective coupling maps, resulting in transpiled circuits that are composed of very different gates despite logically performing the same function. To tackle this issue, our algorithm takes the transpiled baseline circuit and watermarked circuit as inputs, then retrieves the logical circuits by removing the SWAP operations introduced by the quantum backend being used to transpile the circuit. We first iterate through the transpiled circuits (e.g., Fig. \ref{retrieval}) to detect the presence of SWAP gates using the function $is\_swap\_gates$ (lines 1-12), which may manifest themselves in the following ways: (a) direct SWAP gate, explicitly defined as an instance of the 'SwapGate' class, (b) named SWAP gate, which may be defined in a custom way, but carries out the SWAP operation and (c) gate operation for which the unitary matrix is similar to that of a SWAP gate, (d) sequence of 3 CNOT gates with qubit connections which is equivalent to a SWAP gate, (e) iSWAP gate followed by an S gate to reverse the phase from the iSWAP gate, resulting in a SWAP operation, (f) an RXX, RYY, \& RZZ gate each having phase = $pi$/2 which is equivalent to SWAP. Once a SWAP gate is detected (manifesting as (d) in the example shown in Fig. \ref{retrieval}(a) circled with dotted lines), the algorithm reverses them by applying a reverse SWAP gate. We thus get the logical circuits from the transpiled baseline and watermarked circuits, which are now comparable to each other inspite of being transpiled on different backends. 

We then iterate through the logical circuit netlists keeping track of how the gates are configured in the two circuits in $gates1$ \& $gates2$. We run through these lists of tuples, to collect the kinds of gates present and their total count in $count\_gates1$ \& $count\_gates2$ (lines 13-20). In Fig. \ref{retrieval}(a) example, there is one Rz(pi/2) gate on qubit 0, two Rz(pi/4) gates on qubits 1 and 2, and one CNOT gate with control on qubit 1 and target on qubit 0. Similarly, we take the same information for the watermarked circuit. For every gate type in the watermarked circuit, the algorithm checks for differences in the count between the two circuits. The algorithm collects the gate types where this difference is non-zero, in the list $extra\_gates$ (lines 21-28). In our example, this list has one CNOT with control on qubit 1 and target on 0, one Rz(pi/4) gate on qubit 0 and one root X gate. It then traverses through each operation in the watermarked circuit one-by-one, and stores the gate type and the qubit indices for each, as a tuple in $gate\_info$. The counter variable $sequence\_num$ is incremented by 1 for each operation. If this tuple exists in $extra\_gates$, it corresponds to a watermark gate and is added to $watermark\_gates$ (lines 29-38). In our example, we find the watermark gates at sequence numbers 5,6 \& 7. After traversing through the watermarked circuit, we have $watermark\_gates$ that provides the gate type, configuration and sequence number of the watermark constituent gates in the circuit. 
The heuristic has linear time complexity, increasing with the number of watermark gates.

\begin{algorithm}
\caption{Retrieve Watermark}
\label{watermark}
\begin{algorithmic}[1]
\small
\Function {remove\_swaps}{$transpiled\_circuit$}
\State $logical\_circuit \gets \phi$ 

\For{each $instruction$, $qarg$ in $transpiled\_circuit$}
    \If{\Call{is\_swap\_gate}  {instruction}}
        \State $q1, q2 = qarg[0].index, qarg[1].index$
        \State $logical\_circuit \gets (q2, q1)$
    \Else
        \State $logical\_circuit \gets (q1, q2)$
    \EndIf
\EndFor
\State {return} $logical\_circuit$
\EndFunction
\Function {Watermark} {base\_circuit, wm\_circuit}
\State $gates1$ = $Gates( \Call{remove\_swaps}{base\_circuit})$ 
\State $gates2$ = $Gates(\Call{remove\_swaps}{wm\_circuit})$  
\State $count\_gates1$ $\gets$ Type \& frequency of gates in $gates1$
\State $count\_gates2$ $\gets$ Type \& frequency of gates in $gates2$
\State $extra\_gates \gets \phi$ 
\State $watermark\_gates \gets \phi$
\State $ctr$ = 1
\For{each $gate$ in $count\_gates2$}
    \State $diff$ = $count\_gates2[gate]$ - $count\_gates1[gate]$
    \If{$diff$ $>$ 0}
        \For{i in $diff$}
            \State $extra\_gates \gets gate$
        \EndFor
    \EndIf
\EndFor

\For{each $operation$ in \Call{remove\_swaps}{$wm\_circuit$}}
    \State $gate\_info$ = ($gate\_name$, tuple of qubit indices)
    \If{$gate\_info$ in $extra\_gates$}
        \State $watermark\_gates \gets (ctr, gate\_info)$
        \State $extra\_gates.pop(gate\_info)$
    \EndIf
    \State $ctr++$
\EndFor
\State  {return} $watermark\_gates$
\EndFunction
\end{algorithmic}
\end{algorithm}

\subsection{Variations in location and complexity}
Since the watermark gates are placed on the functional as well as auxiliary qubits, we need to choose the location, type and number of gates. To create a good quality watermark, the probability of collision with other legitimate watermarks (quantified using PPA) should be minimized. The PPA calculation is based on the probability of choosing a certain qubit (s) to place the gates. Since each available auxiliary/functional qubit has equal probability of being chosen for placement, the placement of the gate does not impact the PPA.

Adding more gates to the described watermark can make it complex and bring more possibilities while calculating the PPA, reducing it even further. This will reduce the chances of a collision with another watermark. However, it will incur overhead in terms of number of qubits and circuit depth. Making the watermarks complex also affects the time needed by the watermark extraction procedure. We develop and apply a watermark extraction routine to the benchmarks that has been watermarked with a combination of both the models we discussed above, and plot the extraction time. In general, more time is needed to extract the watermark for larger benchmarks.

\begin{figure}
    \centering
        \includegraphics[width=0.65\columnwidth]{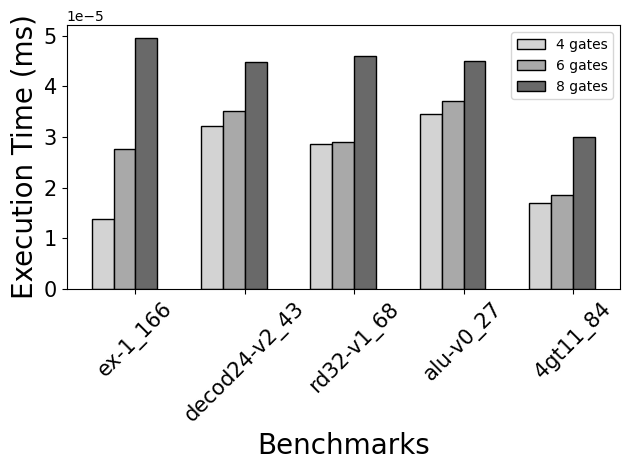}
        \caption{Time needed to extract the watermark.} 
        \label{Qaoa_wm}
     \vspace{-6mm}
\end{figure}

\section{Methodology and Results}
\begin{figure*} [t] 
        \centering         
        \begin{minipage}{0.25\textwidth}
                \centering
                \includegraphics[width=\linewidth]{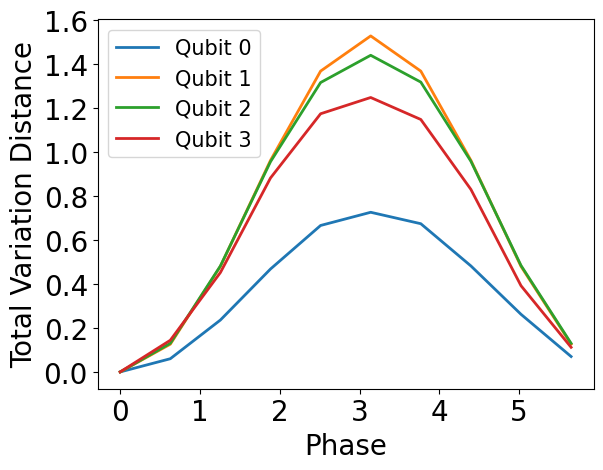}
                 \vspace{-5mm}
                 (a)
        \end{minipage}%
        \begin{minipage}{0.25\textwidth}
                \centering
                \includegraphics[width=\linewidth]{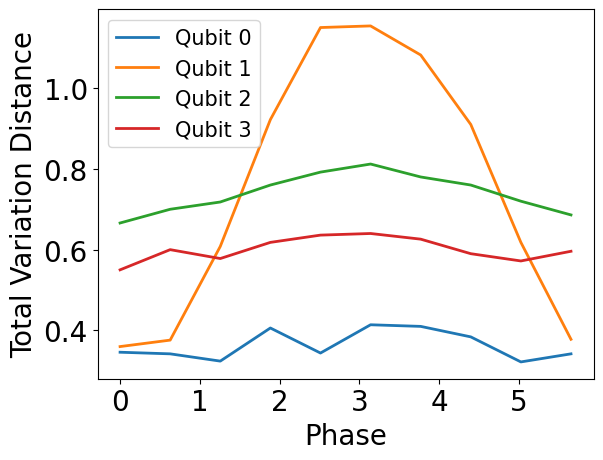}
                 \vspace{-5mm}
                 (b)
        \end{minipage}%
        \begin{minipage}{0.25\textwidth}
                \centering
                \includegraphics[width=\linewidth]{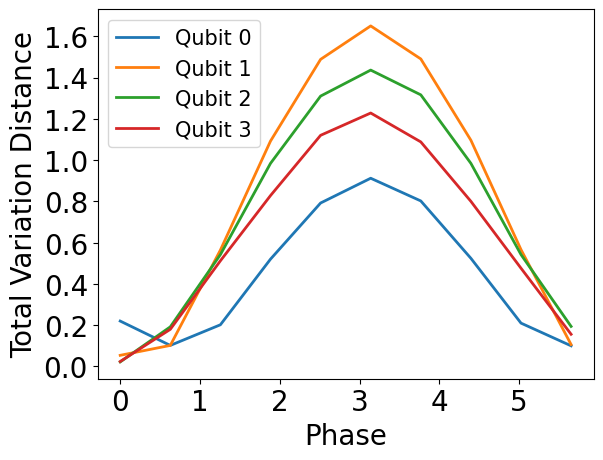}
                 \vspace{-5mm}
                 (c)
        \end{minipage}%
        \begin{minipage}{0.25\textwidth}
                \centering
                \includegraphics[width=\linewidth]{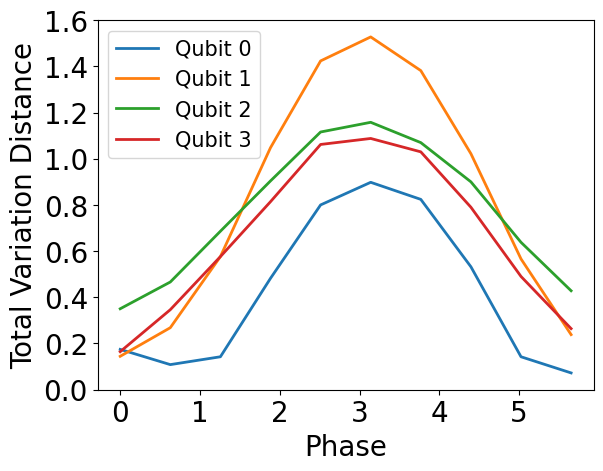}
                 \vspace{-5mm}
                 (d)
        \end{minipage}%
        \vspace{2mm}
        \caption{TVD measured on ancillary qubits for rotation gate (Ry) based watermarking in 4gt4 on (a) FakeValencia (same as base), (b) FakeBogota, (c) FakeKolkata and (d) FakeSherbrooke backends  with respect to original 4gt4 circuit on FakeValencia.}
        \label{benchmarks_TVD}
     \vspace{-4mm}
\end{figure*}

\begin{figure*} [t] 
        \centering         
        \begin{minipage}{0.33\textwidth}
                \centering
                \includegraphics[width=\linewidth]{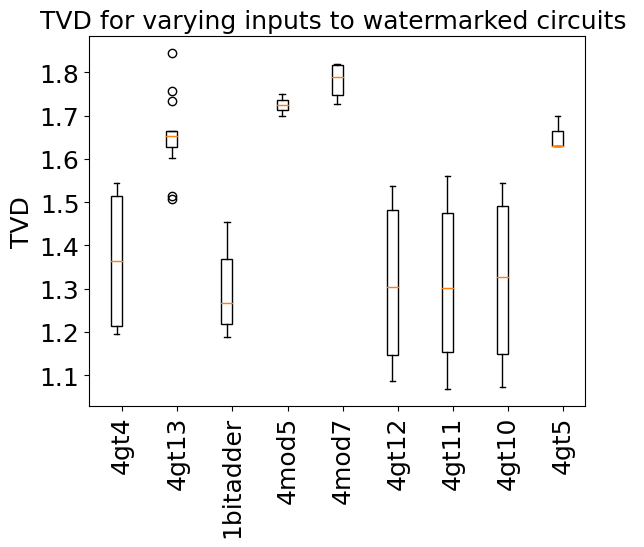}
                 \vspace{-3.5mm}
                 (a)
        \end{minipage}%
        \begin{minipage}{0.33\textwidth}
                \centering
                \includegraphics[width=\linewidth]{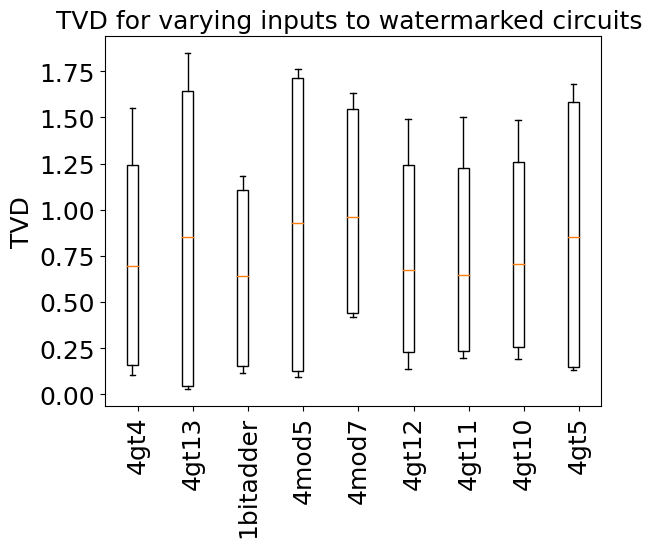}
                 \vspace{-3.5mm}
                 (b)
        \end{minipage}%
        \begin{minipage}{0.33\textwidth}
                \centering
                 \includegraphics[width=\linewidth]{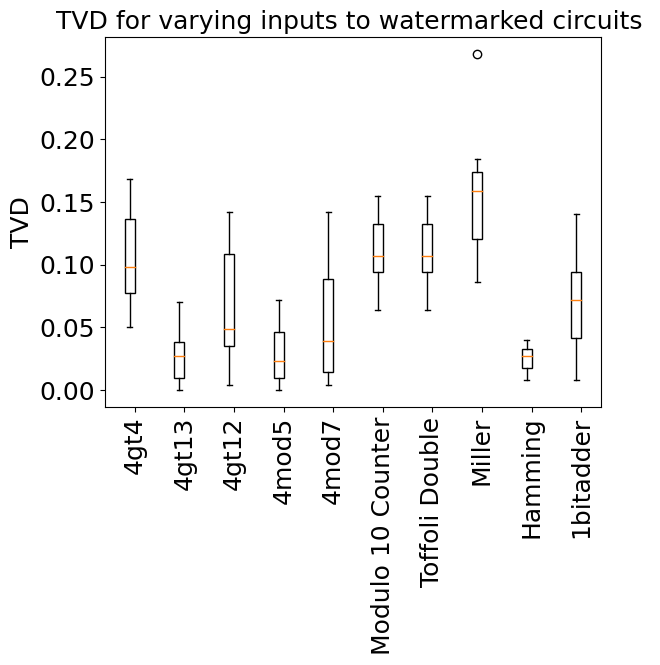}
                 \vspace{-3.5mm}
                 (c)
        \end{minipage}%
        \vspace{-0mm}
        \caption{TVD measured (a) on ancillary qubits for rotation based watermarking for all possible inputs, (b) for rotation \& CNOT based watermarking, (c) on functional qubits for random-gate-set watermarked circuits for all possible inputs.}
        \label{benchmarks_TVD}
        \vspace{-7mm}
\end{figure*}




        


\subsection{Experimental setup and methods}
We run the simulations using IBM Qiskit locally on an AMD Ryzen 5 5500U CPU with Radeon Graphics (2.10 GHz) machine with 8 GB RAM. We took benchmark circuits from the Revlib \cite{b22} repository, which is commonly used for contemporary research on quantum circuit compilation, and generated 3-node, 4-node and 5-node optimized QAOA circuits to apply our watermarks. We use the FakeLagos (7 qubit) and FakeValencia (5-qubit backend) backends of IBM Qiskit for simulations, which use the realistic noise model of ibmq\_lagos (7 qubit) and ibmq\_valencia(5-qubit backend) devices \& FakeBogota (5-qubit backend), FakeSherbrooke (127-qubit backend), FakeKolkata (27-qubit backend) and FakeSingapore backends (20 qubit).

To investigate the effectiveness of the proposed rotation watermark, we run the original 4gt4 circuit and the watermarked circuit for 1000 shots, varying the phase of watermark Ry gate from 0 to 2$\pi$. We then measure the TVD between the probability distribution measured on the ancilla qubit on various backends where the watermark has been placed with respect to the same measurement for the base circuit (Fig. \ref{benchmarks_TVD}). We find that the TVD peaks when an Ry gate of phase $\pi$ is used, averaging at 1.488 across all the backends. 

We then proceed to place an Ry gate of phase $\pi$ on qubit 1. To increase the watermark complexity, we also apply a CNOT gate with qubit 2 acting as control qubit, and qubit 1 acting as target qubit on the benchmark circuits where these qubit positions are ancillary. The TVDs are measured between the probability distribution measured on the ancilla qubit where the watermark has been placed with respect to the same measurement for the base circuit for all possible inputs (Fig. \ref{benchmarks_TVD}(b)). We note that the TVD for each circuit over all inputs is significant, averaging at 0.816 \& going up to 0.994.

To apply this watermark design to a QAOA circuit, we add an ancillary qubit to the QAOA maxcut circuits generated for a graph inputs since all qubits are functional. We then measure the approximation ratios of the optimized 3-node, 4-node \& 5-node QAOA graph maxcut circuits for 2 layers \& 3 layers with and without the watermark gates i.e.,  CNOT gate placed on the qubit 2 (control) \& the ancilla qubit (target) and an Rx gate with phase pi on qubit 1 (Table \ref{tab:approximation-ratios}). For the 3 \& 4-node QAOA circuits we use FakeValencia backend, but for 5 nodes we use FakeSingapore backend to overcome the qubit limitation. We note a small drop in approximation ratio, upto 0.051, on applying the watermark. In some cases we note no impact or even improved approximation ratio after adding random gate based watermark. This can be explained by the fact that in these cases the additional random gates increase the complexity of the quantum state space providing more opportunities to find an optimal solution \cite{b27}. \\
For the random gate based watermarking, we apply a Pauli-X gate on qubit 0 and a CNOT gate with qubit 2 as control and qubit 1 as target qubit followed by their inverse gates, in the middle of the benchmarking circuits. The watermarking gate set is separated by a barrier with the inverse gate set. We then measure the TVD between the probability distribution measured on the functional qubit (s) with respect to the same measurement for the base circuit for all possible inputs (Fig. \ref{benchmarks_TVD}(c)). We observe an average TVD of 0.07061, peaking at 0.1565 over all the circuits and inputs.
\begin{figure}
    \centering
        \includegraphics[width=0.85\columnwidth]{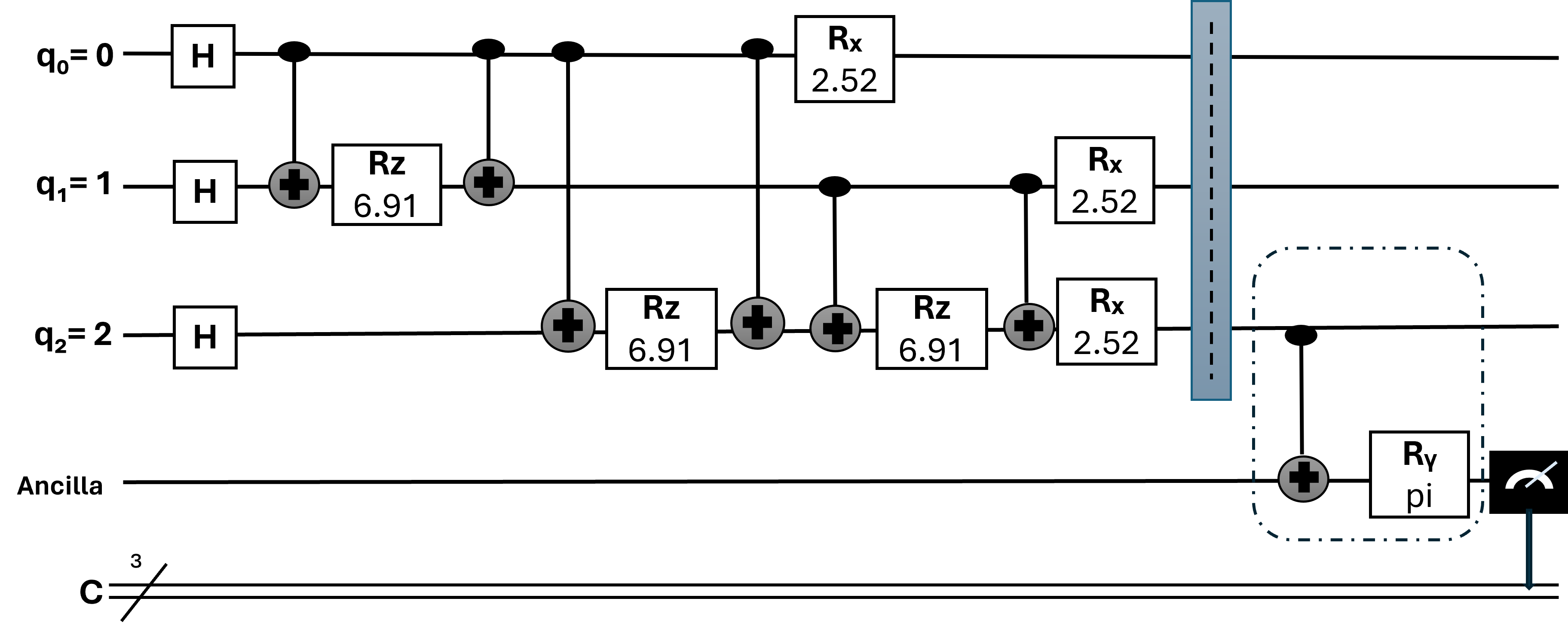}
         \vspace{-2mm}
        \caption{3-node QAOA circuit with extra ancilla qubit to insert rotation gate based watermarking (encircled).}
        \vspace{-9mm}

        \label{Qaoa_wm}
\end{figure}

\begin{table*}
\centering
\vspace{2mm}
\caption{Approximation ratios for various QAOA circuits with and without rotation and random gate based watermarking.}
\resizebox{\textwidth}{!}{
\begin{tabular}{|c|c|c|c|c|c|c|c|c|}
\hline
& \multicolumn{4}{c}{Rotation Gate Based Watermarking} & \multicolumn{4}{|c|}{Random Gate Based Watermarking}\\
\cline{2-9}
Nodes & \multicolumn{2}{c|}{2 layers} & \multicolumn{2}{c|}{3 layers} & \multicolumn{2}{c|}{2 layers} & \multicolumn{2}{c|}{3 layers} \\
\cline{2-9}
& W/o watermark & W/ watermark & W/o watermark & W/ watermark & W/o watermark & W/ watermark & W/o watermark & W/ watermark  \\
\hline
3-node & 0.939 & 0.926 & 0.929 & 0.893 & 0.765 & 0.773 & 0.777 & 0.923 \\\hline
4-node & 0.807 & 0.756 & 0.780 & 0.748 & 0.770 & 0.761 & 0.794 & 0.777 \\\hline
5-node & 0.688 & 0.688 & 0.688 & 0.688 & 0.701 & 0.708 & 0.719 & 0.703\\
\hline
\end{tabular}}
\label{tab:approximation-ratios}
\end{table*}


\begin{table*}
\centering
\caption{The comparison between our multi-stage watermarking technique and the state-of-the-art.}
\resizebox{\textwidth}{!}{
\small
\begin{tabular}{|c|c|c|c|c|c|c|c|c|c|c|c|c|c|c|c|}
\hline
\multirow{2}{*}{Benchmarks} & \multicolumn{3}{c|}{non-watermarked} & \multicolumn{4}{c|}{Ours} & \multicolumn{4}{c|}{Improvement w.r.t. \cite{b24}} & \multicolumn{4}{c|}{Improvement w.r.t. \cite{b26}}\\\cline{2-16}
& \makecell{Circuit \\ Depth} & \makecell{2-qubit \\ Gate \#}  & PST & \makecell{Circuit \\ Depth} & \makecell{2-qubit \\ Gate \#}  & PST & PPA & \makecell{Circuit \\ Depth(\%)} & \makecell{2-qubit \\ Gate \#(\%)}  & PST(\%) & PPA(x) & \makecell{Circuit \\ Depth(\%)} & \makecell{2-qubit \\ Gate \#(\%)}  & PST(\%) & PPA(x)\\
\hline
ex-1\_166 & 27 & 19 & 0.875 & 31 & 23 & 0.875 & 1.5E-6 & \textbf{16.13} & \textbf{21.74} & \textbf{3.98} & \textbf{23800x} & \textbf{0} & -30.43 & \textbf{5.7} & \textbf{166667x}\\\hline
decod24-v2\_43 & 70 & 52 & 0.723  & 77 & 58 & 0.724 & 2.4E-7 & -18.18 & -44.83 & -9.99 & \textbf{86667x} & -10.39 & -39.66 & \textbf{1.24} & \textbf{260416x}\\\hline
rd32-v1\_68 & 43 & 31 & 0.889  & 52 & 39 & 0.89 & 2.4E-7 & \textbf{1.92} & -33.33 & \textbf{11.42} & \textbf{43750x} & \textbf{11.54} & \textbf{23.08} & \textbf{15.47} & \textbf{260416x}\\\hline
alu-v0\_27 & 46 & 37 & 0.932 & 63 & 48 & 0.909 & 2.4E-7 & \textbf{0} & -27.08 & \textbf{19.23} & \textbf{148750x} & \textbf{17.46} & \textbf{10.42} & \textbf{42.45} & \textbf{65000x} \\\hline
4gt11\_84 & 23 & 16 & 0.964 & 25 & 19 & 0.96 & 2.4E-7 & \textbf{52} & -10.53 & \textbf{12.78} & \textbf{238333x} & \textbf{28} & -10.53 & \textbf{15.65} & \textbf{1041666x} \\\hline
\end{tabular}}
\label{tab:comparison}
\end{table*}

\section{Security and Comparative Analysis}

\subsection{Comparison with state-of-the-art}

For comparative analysis, we combine both watermark models introduced in this paper and compare various metrics such as the circuit depth, 2-qubit gate count for transpiled circuits, PST and PPA against the non-watermarked and the watermarked circuits in \cite{b24} and \cite{b26}. To ensure uniformity, we execute the benchmark circuits on FakeLagos backend as used to extract results for \cite{b24} \& \cite{b26}, and setting optimization\_level=2 for transpilation. 

To elaborate PPA calculation, let us consider an example. Qiskit has a total of 10 single qubit and 12 2-qubit gates. Since the random gate based watermark in our experiments uses 2 kinds of gates (NOT \& CNOT), there are $\binom{22}{2}$ = 231 possible selections. We then calculate the possible placements of the chosen gates over the number of qubits available in the circuit (considering the number of qubits in the circuit and the ancillary qubit that may be added to apply the rotation gate based watermark). In addition to the random gates, we also have a rotation gate based watermark consisting of an Rx/Ry gate  (which could be any one of the 10 single qubit gates available in Qiskit, so 10 possibilities) in association with a CNOT gate. Having chosen a rotation gate, we consider the possible phase angles that may be taken for the watermark. Considering a resolution of $\pi$/6, there could be 6 possibilities. We multiply all the possibilities to know the total number of possible watermarks that could be generated. For example, ex-1\_166 benchmark can give rise to 231 $\times$ 4 $\times$ 12 $\times$ 10 $\times$ 6 = 249480 possibilities for the combined watermark. We note that the PPA for our watermarks is astronomically smaller in comparison to the state-of-the-art.

The circuit depth for our watermark shows an average 18.28\% overhead. With respect to \cite{b24}, we have reduced circuit depth by about 7.1\%, and by 27.7\% with respect to \cite{b26}. The PST reduced only by a miniscule 0.53\% on an average due to the proposed watermark. Compared to \cite{b24}, our PST is 22.69\% higher and 9.28\% more as against \cite{b26} on average. However, we note an overhead of 21.38\% for the 2-qubit gate count with respect to the non-watermarked circuits. An added 2-qubit gate count overhead of 28.36\% and 41.59\% with respect to \cite{b26} and \cite{b24}, respectively. 
\vspace{-0.15cm}

\subsection{Watermark tampering \& removal}

\vspace{-0.1cm}

Tampering or removal of watermark by an adversary can prevent the designer from claiming legitimate ownership of the quantum circuit. In the random gate based watermark, we use a barrier to separate the watermarking gates from its inverse. However, the barrier is be visible post transpilation, making the watermark resilient against identification and removal. For the rotation gate based watermarking in the auxiliary qubits, the presence of a rotation gate or an extra ancillary qubit in the quantum implementation of an arithmetic circuit (such as 4gt4) may raise suspicion, since a rotation gate is usually not expected in such circuits. The adversary may identify and remove the rotation gates and/or ancillary qubits. To defend against this attack, we propose to apply multiple watermarks to a circuit so that the removal of one watermark does not lead to obliteration of the maker's signature. 
\subsection{Ghost watermark}
An adversary may introduce a ghost signature in the cloned circuit i.e., they will announce the presence of a watermark even when there isn't one, to generate false ownership of the stolen design. 
We can solve this problem by applying multiple watermarks 
making it hard for adversary to claim the circuit. 
\vspace{-0cm}
\section{Conclusions}
We present two lightweight watermarking techniques to help the designer prove ownership in the event of an adversarial theft by third party quantum cloud services. 
The overhead of the proposed watermarks and various other metrics are found to be considerably better than the existing techniques. 

\section{Acknowledgment}
This work is supported in parts by NSF (CNS-1722557, CNS-2129675, CCF-2210963, CCF-1718474, OIA-2040667, DGE-1723687, DGE-1821766 and DGE-2113839) and Intel’s gift.


\section*{}

\vspace{12pt}

\end{document}